%Paper: cond-mat/9208024
%From: YSHA@db1.cc.rochester.edu
%Date: Sat, 29 Aug 1992 14:14 EDT

%--------------------------------%
%  GENERALLY USEFUL DEFINITIONS  %
%--------------------------------%

\def\today{\number\day\enspace
     \ifcase\month\or January\or February\or March\or April\or May\or
     June\or July\or August\or September\or October\or
     November\or December\fi \enspace\number\year}
\def\clock{\count0=\time \divide\count0 by 60
    \count1=\count0 \multiply\count1 by -60 \advance\count1 by \time
    \number\count0:\ifnum\count1<10{0\number\count1}\else\number\count1\fi}
\footline={\ifnum\pageno=1 \hss \hfill \hss
           \else \hss\rm -- \folio\ -- \hss\fi}
% will show up on top of each page of draft, if redefined

\overfullrule=0pt % Gets rid of the black box next to "overfull" lines
\raggedbottom
\def\push#1{\hbox to 1truein{\hfill #1}}

%-----------%
%  SYMBOLS  %
%-----------%

\def\arcsec{\ifmmode^{\prime\prime}\else $^{\prime\prime}$\fi}
\def\arcmin{\ifmmode^{\prime}\else $^{\prime}$\fi}
\def\deg{\ifmmode^\circ\else$^\circ$\fi}

\def\Teff{{\it T}\lower.5ex\hbox{\rm eff}}
\def\kms{\ifmmode{{\rm km~s}^{-1}}\else{km~s$^{-1}$}\fi}
\def\pervol{\ifmmode{{\rm cm}^{-3}}\else{cm$^{-3}$}\fi}
\def\perarea{\ifmmode{{\rm cm}^{-2}}\else{cm$^{-2}$}\fi}
\def\mum{\ifmmode{\mu {\rm m}}\else{$\mu$m}\fi}
\def\respwr{\ifmmode{\lambda /\Delta\lambda}
    \else{$\lambda /\Delta\lambda$}\fi}
\def\whz{\ifmmode{{\rm W~Hz}^{-1/2}}\else{W~Hz$^{-1/2}$}\fi}
\def\wig#1{\mathrel{\hbox{\hbox to 0pt{%
    \lower.5ex\hbox{$\sim$}\hss}\raise.4ex\hbox{$#1$}}}}

\def\sqr#1#2{{\vcenter{\vbox{\hrule height.#2pt
    \hbox{\vrule width.#2pt height#1pt \kern#1pt
    \vrule width.#2pt}
    \hrule height.#2pt}}}}
\def\square{\mathchoice\sqr34\sqr34\sqr{2.1}3\sqr{1.5}3}
\def\boxit#1{\vbox{\hrule\hbox{\vrule\kern3pt
   \vbox{\kern3pt#1\kern3pt}\kern3pt\vrule}\hrule}}

                %if you want a space after these,
              %you must put a hard space in.
\def\h{{\rm h}}

\def\ie{{\it i.e.\/}}

\hyphenation{nucleo-cos-mo-chron-ology}  % Hypenations that could be useful
\hyphenation{Nucleo-cos-mo-chron-ology}
\hyphenation{chron-o-meter}
\hyphenation{Ro-ches-ter}
\hyphenation{mol-e-cules}

%--------------%
%  REFERENCES  %
%--------------%

\def\jref#1 #2 #3 #4 {{\par\noindent \hangindent=3em \hangafter=1
      \advance \rightskip by 5em #1, {\it#2}, {\bf#3}, #4.\par}}
\def\ref#1{{\par\noindent \hangindent=3em \hangafter=1
      \advance \rightskip by 5em #1\par}}

%----------------------------------%
%  INCREMENTAL EQUATION NUMBERING  %
%----------------------------------%

\newcount\eqnum
\def\nexteq{\global\advance\eqnum by1 \eqno(\number\eqnum)}
\def\lasteq#1{\if)#1[\number\eqnum]\else(\number\eqnum)\fi#1}
\def\preveq#1#2{{\advance\eqnum by-#1
    \if)#2[\number\eqnum]\else(\number\eqnum)\fi}#2}

%----------%
%  TABLES  %
%----------%

\def\tableheight{\vrule width 0pt height 8.5pt depth 3.5pt}
{\catcode`|=\active \catcode`&=\active
    \gdef\tabledelim{\catcode`|=\active \let|=\vbar
                     \catcode`&=\active \let&=\nobar} }
\def\table{\begingroup
    \def\twidth{\hsize}
    \def\tablewidth##1{\def\twidth{##1}}
    \def\defaultheight{\vrule width 0pt height 8.5pt depth 3.5pt}
    \def\heightdepth##1{\dimen0=##1
        \ifdim\dimen0>5pt
            \divide\dimen0 by 2 \advance\dimen0 by 2.5pt
            \dimen1=\dimen0 \advance\dimen1 by -5pt
            \vrule width 0pt height \the\dimen0  depth \the\dimen1
        \else  \divide\dimen0 by 2
            \vrule width 0pt height \the\dimen0  depth \the\dimen0 \fi}
    \def\spacing##1{\def\defaultheight{\heightdepth{##1}}}
    \def\nextheight##1{\noalign{\gdef\tableheight{\heightdepth{##1}}}}
    \def\end{\cr\noalign{\gdef\tableheight{\defaultheight}}}
    \def\zerowidth##1{\omit\hidewidth ##1 \hidewidth}
    \def\hline{\noalign{\hrule}}
    \def\skip##1{\noalign{\vskip##1}}
    \def\bskip##1{\noalign{\hbox to \twidth{\vrule height##1 depth 0pt \hfil
        \vrule height##1 depth 0pt}}}
    \def\header##1{\noalign{\hbox to \twidth{\hfil ##1 \unskip\hfil}}}
    \def\bheader##1{\noalign{\hbox to \twidth{\vrule\hfil ##1
        \unskip\hfil\vrule}}}
    \def\spanloop{\span\omit \advance\mscount by -1}
    \def\extend##1##2{\omit
        \mscount=##1 \multiply\mscount by 2 \advance\mscount by -1
        \loop\ifnum\mscount>1 \spanloop\repeat \ \hfil ##2 \unskip\hfil}
    \def\vbar{&\vrule&}
    \def\nobar{&&}
    \def\hdash##1{ \noalign{ \relax \gdef\tableheight{\heightdepth{0pt}}
        \toks0={} \count0=1 \count1=0 \putout##1\end
        \toks0=\expandafter{\the\toks0 &\end} \xdef\piggy{\the\toks0} }
        \piggy}
    \let\e=\expandafter
    \def\putspace{\ifnum\count0>1 \advance\count0 by -1
        \toks0=\e\e\e{\the\e\toks0\e&\e\multispan\e{\the\count0}\hfill}
        \fi \count0=0 }
    \def\putrule{\ifnum\count1>0 \advance\count1 by 1

\toks0=\e\e\e{\the\e\toks0\e&\e\multispan\e{\the\count1}\leaders\hrule\h
fill}
        \fi \count1=0 }
    \def\putout##1{\ifx##1\end \putspace \putrule \let\next=\relax
        \else \let\next=\putout
            \ifx##1- \advance\count1 by 2 \putspace
            \else    \advance\count0 by 2 \putrule \fi \fi \next}   }
\def\tablespec#1{
    \def\vdimens{\noexpand\tableheight}
    \def\tabby{\tabskip=0pt plus100pt minus100pt}
    \def\r{&################\tabby&\hfil################\unskip}
    \def\c{&################\tabby&\hfil################\unskip\hfil}
    \def\l{&################\tabby&################\unskip\hfil}
    \edef\templ{\noexpand\vdimens ########\unskip  #1
         \unskip&########\tabskip=0pt&########\cr}
    \tabledelim
    \edef\body##1{ \vbox{
        \tabskip=0pt \offinterlineskip
        \halign to \twidth {\templ ##1}}}
% Added by MJWengler 86/8/7, allows pagebreaking in middle of table by
% not vboxing it, no idea what you lose by doing this.
    \edef\sbody##1{ {
        \tabskip=0pt \offinterlineskip
        \halign to \twidth {\templ ##1}}}
}% end of \table definition

%-----------%
%  FORMATS  %
%-----------%

\def\ltextindent#1{\indent\llap{\hbox to \parindent {#1\hfil}}\ignorespaces}

\def\mathfont#1{{
    #1\count20=\fam\multiply\count20 by "100\advance\count20 by "7000
    %\count20 we will make mathcodes for letter x = x+\count20
    \count21=`a \advance\count21 by - 1
    \count22=\count21\advance\count22 by \count20
    \loop \advance\count22 by 1 \advance\count21 by 1
             \global\mathcode\count21=\count22
    \ifnum \count21<`z \repeat
    \count21=`A \advance\count21 by - 1
    \count22=\count21\advance\count22 by \count20
    \loop \advance\count22 by 1 \advance\count21 by 1
             \global\mathcode\count21=\count22
    \ifnum \count21<`Z \repeat}}

\outer\def\section#1\par{\vskip 12pt plus 10pt minus 4pt
    \centerline{\bf#1}\nobreak\vskip 5pt plus 3pt minus 2pt}

%-----------------------%
%   LASERWRITER FONTS   %
%-----------------------%

%\font\rm=cmtimer
%\font\it=cmtimei
%\font\bf=cmtimeb
\font\title=cmbx10 scaled \magstep2
%\rm

%\input[admin.mkfa.texd]abcmacro.tex
\magnification=1200
\baselineskip=20pt
\def\Int{{\rm Int}}

\centerline{\title Localized Electrons on a Lattice}
\centerline{\title with Incommensurate Magnetic Flux}
\vskip 10pt
\centerline{\it Shmuel Fishman}
\centerline{Department of Physics}
\centerline{Technion, IIT}
\centerline{32000 Haifa, Israel}
\vskip 10pt

\centerline{\it Yonathan Shapir}
\centerline{Department of Physics and Astronomy}
\centerline{University of Rochester}
\centerline{Rochester, NY \ 14627-0011}
\centerline{USA}
\vskip 10pt

\centerline{\it Xiang-Rong Wang}
\centerline{School of Physics and Astronomy}
\centerline{University of Minnesota}
\centerline{Minneapolis, MN \ 55455}
\centerline{USA}
%\vskip 5pt
\centerline{and}
%\vskip 5pt
\centerline{Department of Physics}
\centerline{The Hong Kong University}
\centerline{of Science and Technology}
\centerline{Clear Water Bay, Kowloon}
\centerline{Hong Kong}
\vfill\eject

\centerline{\bf ABSTRACT}
\vskip 10pt

The magnetic field effects on lattice wavefunctions of Hofstadter electrons
strongly localized at boundaries are
studied analytically and numerically.
The exponential decay of the wavefunction is modulated by a field dependent
amplitude $J(t)=\Pi_{r=0}^{t-1} 2\cos(\pi\alpha r)$, where
$\alpha$ is the magnetic flux per plaquette (in units of a flux quantum)
and $t$ is
the distance from the boundary (in units of the lattice spacing).  The behavior
of $|J(t)|$ is found to depend sensitively on the value of $\alpha$. \
While for rational values $\alpha=p/q$, the envelope of $J(t)$ increases as
$2^{t/q}$, the behavior for $\alpha$ irrational $(q\to\infty)$ is erratic
with an aperiodic structure which drastically changes with $\alpha$. \
For algebraic $\alpha$ it is found that $J(t)$ increases as a power law
$t^{\beta(\alpha)}$ while it grows faster (presumably as $t^{\beta(\alpha)\
ln\ t}$) for transcendental $\alpha$.
This is very different from the growth rate $J(t)\sim e^{\sqrt{t}}$ that is
typical for cosines with random phases.  The theoretical analysis is
extended to products of the type $J^\nu(t)=\Pi_{r=0}^{t-1}
2\cos(\pi\alpha r^\nu)$ with $\nu>0$. \ Different behavior of $J^\nu (t)$ is
fou
nd
in various regimes of $\nu$. \ It changes from periodic for small $\nu$
to random like for large $\nu$.

\vfill\eject

\noindent {\bf I. \ Introduction}

The properties of non-interacting electrons on a lattice subjected to an
external magnetic field have attracted much attention since the early works
of Hofstadter,$^1$ Wannier,$^2$ and Azbel.$^3$ \ These works have primarily
focused on the exotic spectral properties as a function of the electron's
energy and the parameter $\alpha =\phi/\phi_0$ where $\phi$ is the
magnetic flux per plaquette and $\phi_0=\hbar/e$ is the flux quantum.  The
spectrum has special scaling properties as a function of the
commensurability $q$ for $\alpha=p/q$ (rational) and becomes a Cantor set
for incommensurate fluxes ($\alpha$ irrational).
The wavefunction itself is extended in the $q$ subbands for commensurate
values of rational $\alpha$. \

The Hamiltonian describing the lattice
electron is:
$${\cal H} =W\sum_i a_i^\dagger a_i +V\sum_{\langle
ij\rangle}a_i^\dagger a_j e^{i\gamma_{ij}} +c.c.\eqno(1.1)$$
The phase $\gamma_{ij}$ is associated to the link $\langle ij\rangle$
between the nearest-neighbor sites in accordance with the ``Peierls
ansatz".  Any gauge such that the sum around a plaquette
$$\sum_{\square} \gamma_{ij}=e\phi/\hbar\eqno(1.2)$$
will do.

More recently$^{4-7}$ there is growing interest in the effect of magnetic
field on localized electrons.  The combined effects of lattice periodicity and
magnetic flux create a very complex behavior in the spatial variation of the
wavefunctions even in the absence of disorder.$^4$ \ This is the case when the
electron's energy is deep inside a gap between the quasi-bands of the bulk
eigenstates.$^{1-3}$ \ The electron may be localized at inhomogeneities such
as the edge of the lattice (\ie\ surface gap states) or at isolated impurities
in an otherwise ordered bulk.  Clearly these states
decay exponentially going away from the inhomogeneity into the bulk.  This
exponential decay and particularly the ``localization" length (associated
with this exponential decay) will be affected by the application of an
external magnetic field to the system.

In the present work we look at some basic aspects of this problem.  While
avoiding the full complexities of the related questions, we look at a simple
(maybe the simpler) model in which the intricate salient features of this
problem are predominant and can be addressed.

As it is well known$^{1-3}$
the 2D tight-binding lattice electron problem reduces
(in the Landau gauge) to that of a 1D electron hopping in a potential which
varies as \break $\cos(k_\perp j+2\pi\alpha)$ where $k_\perp$ are
the transverse momenta in the $y$ direction, and may be taken to be zero by
shifting the origin in the $x$ direction.$^1$ \ The Schr\"odinger
equation reduces then to the famous Harper equation:
$$u_{n+1}+u_{n-1}+V_nu_n=(E-W)u_n\eqno(1.3)$$
where the diagonal potential is
$$V_n=\lambda\cos\pi\alpha n.\eqno(1.4)$$

As discussed below this model was generalized$^{8-10}$ to potentials of the
type:
$$V_n^\nu=\lambda\cos\pi\alpha n^\nu.\eqno(1.5)$$
It can be used in order to analyze the transition from extended
to localized wavefunctions in the proposed experimentally
realizable layered systems$^{10,11}$ in which the distance between adjacent
layers increases as $n^\nu$ (for $0<\nu<1)$. \ The quantum transmission
of such layered systems in the strong localization regime, for different
values of $\nu$, is also a part of our present investigations.

In the localized regime, the decay of the wavefunction may be studied by
looking
at the probability of an electron localized at the origin to tunnel to another
site a distance $t$ away, $|I(t)|^2$, where $I(t)=\langle \psi^*_0(t)\psi_0(0)
\rangle$ is the related Green's function $(\psi_0(r)$ is the wavefunction
localized at the origin).  Our results will be derived within the ``directed
paths" approximation$^7$ which becomes better as the localization (=decay)
length becomes smaller.  This will be the case when the hopping matrix elements
V is much smaller than the on site energy W and the energy
$E$ is far away from the band $W\pm2V$ of the extended bulk eigenstates
($E=0$ is a convenient choice which fulfills this requirement).

Preliminary results of investigations which go beyond this approximation
and include ``returning loops," are reported in the last chapter.

The Green's function $G(r)$ may then be expressed as a sum over paths and
since each step has an amplitude of $V/W$ (for $E=0$), the leading contribution
comes from the shortest, hence directed, paths.  If only these are kept
one has $^4$:
$$|I(t)|^2 = \left({V\over W}\right)^{2t}|J(t)|^2\eqno(1.6)$$
Where $(V/W)^{2t}$ is responsible for the strong exponential decay
and $J(t)$ contains all the interference effects.  In absence of magnetic
field $J(t)$ is just the total number of paths going from the origin to the
final site which increases as $2^{2t}$ (note that since $V<<2W$, $I(t)$ still
strongly decays exponentially with $t$).  In presence of magnetic field
$J(t)$ has a very complex behavior as function of the flux $\alpha$ and the
distance $t$. \ $J(t)$ also depends explicitly on the geometry and the one
utilized in most of the works is that of a square lattice with the origin and
final sites being along the diagonal (this choice is a natural one; similar
calculations, however, may be carried for any locations of these sites).  For
the surface realizations this choice means that the edge is in the
$[\overline1, 1]$ direction (the distance along this direction will hereafter
be denoted $x$), while the direction perpendicular to the edge is
$[1,1]$ (and the distance from the surface into the bulk, in this direction
is $t$).

The $J(x,t)$ for consecutive $t$ are related by a transfer matrix $T$:
$$J(x,t+1)=\sum_{x^\prime}T_{t,t+1}(x,x^\prime)J(x^\prime,t)\eqno(1.7)$$

The solution relies on the diagonalization of $T$ and the technical
calculations are given elsewhere $^4$, and will not be repeated here.  The
impor
tant
feature is that in presence of magnetic fields a gauge may be chosen
such that the matrices which will depend explicitly on $t$ will, nevertheless,
commute and therefore may be diagonalizable simultaneously.  The
eigenvectors are the transverse waves $e^{ik_\perp x}$ with $k_\perp=\pm
m\pi/L$ ($L>>t$ is a large width cutoff).  So the localized state at
$(0,0)$ may be decomposed into transverse Fourier components ($x$ direction)
and each component (with $k_\perp)$ will ``propagate" independently into the
bulk in the $t$ direction.  It is thus very natural to define the quantity
$J(t,k_\perp)$ for each Fourier component.  In previous calculations it was
found for given $k_\perp$ and $\alpha=\phi/\phi_0$:
$$|J_\alpha(t,k_\perp)|^2=\Pi_{r=0}^{t-1} |2\cos(\pi\alpha
r-k_\perp)|^2.\eqno(1.8)$$

The asymptotic behavior of these products do determine  the decay into the
bulk of the surface gap states which in the transverse direction along the
surface have the same dependence as the eigenvectors, \ie\ $e^{ik_\perp x}$. \

To find $G(x,t)$ for an electron localized at a site we still
need to sum factors like $J_\alpha(k_\perp,t)e^{ik_\perp x}$ over all
$k_\perp$.
The asymptotic behavior in real space will be determined by that of $J_\alpha
(k_\perp,t)$ for large $t$ as we observed numerically.  The closed form of
$G(x,t)$ requires the knowledge of the exact amplitude and phase of all
$J_\alpha(k_\perp,t)$ and is beyond the scope of the present paper.

The behavior of products as in Eq.~(1.8) is very sensitive to the values of
$\al
pha$.
\ For any rational $\alpha=p/q$ it may be shown that $|J(t)|$ increases
as $2^{t/q}$. \ That naturally raises the question what will be the
behavior as $q\to\infty$. \ This and related questions are the subject
of the present paper.

Preliminary numerical investigation for $\alpha=(\sqrt 5 -1)/2$ (the
golden mean) have exhibited $J_\alpha(x,t)$ in the form of bounded aperiodic
fluctuations. \ Although an exponential behavior where
$q$ is the scale is ruled out, other possibilities like powers of $t$
of $\ln t$ are still possible.

Another asymptotic behavior we should consider arises if the phases of the
cosines in the product are random.  Then the behavior is that of a product
of random variables.  The typical (though not the average) behavior is
$e^{\langle\log J\rangle}\sim e^{c\sqrt t}$.  It should be noted that this
behavior corresponds, in the original lattice model, to a magnetic flux which
is uniform in the $x$ direction but changes randomly from one row to the
next in the $t$ direction (so-called ``random rods").

The initial motivation for the generalization of the potential
$V_n^\nu$ (Eq. (1.5)) to $\nu\not=1$ came from the field of quantum chaos.  The
kicked rotor model was mapped on the tight-binding model
with the diagonal potential$^{12}$
$$\tilde V_n^\nu=\tan\pi\alpha n^2.\eqno(1.9)$$
It was argued that it behaves like a random potential, leading to localization
of the eigenstates of the corresponding model.  In the field
of quantum chaos it explains the quantal suppression of chaos.
It was argued that the sequence (1.9) is pseudorandom since if $n$ changes
by 1, the phase changes by $2\pi\alpha n$ that is a large number.  Since
the tangent depends only on the phase mod$\pi$ it depends on a small fraction
of a large number.  Therefore, if $\alpha$ is irrational it is related to
remote digits in a representation of an irrational number and therefore can be
considered pseudorandom.  The theoretical investigation of this issue
motivated the introduction$^8$ of the tight-binding model (1.3) with the
potential (1.5). \ The parameter $\nu$ controls the degree of pseudorandomness,
that increases with $\nu$. \ It was found that the asymptotic behavior of sums
of the form
$$S_N^\nu=\sum_{n=1}^N V_n^\nu\eqno(1.10)$$
is of crucial importance for the understanding of localization.  Sums of
this form$^{13-15}$ are of great importance for the present work as well.
The pseudorandom properties of the sequences (1.5) were classified in the
framework of standard tests for pseudorandomness.$^{16}$ \
One of the conclusions of these investigations is that they are very different
for various regimes of $\nu$. \ For $\nu>2$ the behavior of the sequences
is very similar to the one of the corresponding random ones, namely to
sequences where the phase is truly random.  For $1<\nu<2$ the asymptotic
growth of the sums (1.10) is $S_N^\nu\sim\sqrt{N}$, namely as for
random ones, but the growth takes place$^{8,9}$ for narrow regions in $n$. \
For $0<\nu<1$ the difference between consecutive terms approaches zero
for large $n$ and the sequences do not resemble at all random sequences.
The physically important cases $\nu=1$ and $\nu=2$ are bordering cases between
different regimes.  Although the generalization (1.5) of the usual Harper
equation was proposed$^{8-10}$ for purely theoretical reasons, it was
suggested to be relevant for plasmon dynamics in artificially constructed
superlattices.$^{10-11}$ \ Effects of such potentials on modulated
waveguides in the microwave regime may be investigated as well.$^{17}$

In the present paper the asymptotic behavior of $J(t)$ is investigated.  For
this purpose the sum
$$A_N=\ln|J(t=N)|\eqno(1.11)$$
is investigated in Sec.~II. \ It turns out to be very different from
the corresponding quantity where the phases are random.

It can be generalized to arbitrary $\nu$ as
$$A_N^\nu=\ln |J^\nu(t=N)|=\ln [\Pi_{r=0}^{N-1}2|\cos\pi\alpha r^\nu|]
\eqno(1.12)$$
In the end of Sec.~II it is shown that a small perturbation in $\nu$ around 1
yields a drastic change in the behavior of $A_N^\nu$. \ In Sec.~III the
behavior of the sums (1.10) is studied in the regimes $0<\nu<{1\over 2}$,
${1\over2}<\nu<1$, $1<\nu<2$ and $\nu\ge2$. \ In each of these regimes
one finds different behavior that differs also from the one found for $\nu=1$.
\

The results are summarized in Sec.~ IV.
\vskip 10pt

\noindent{\bf II. \ The Sums $A_N$ for the Incommensurate Phase
($\nu=1$)}
\vskip 10pt

In this section we will estimate the sum of (1.11) namely,
$$A_N=\sum_{n=0}^{N-1}\ln 2|\cos(\pi\phi_n)|\eqno(2.1)$$
with
$$\phi_n=\alpha n\eqno(2.2)$$
Of particular interest will be the difference between this sum and the
corresponding random sum, namely the sum where the phase $\phi_n$ is
replaced by a random variable that is uniformly distributed in the interval
$[-1,1]$. \ In order to estimate the sum (2.1) we will exploit the fact
that each term is periodic in $\phi_n$. \ It is easy to see that
$$\ln|\sin\theta|=-\ln 2-\sum_{m=1}^\infty {1\over m}\cos
2m\theta\eqno(2.3)$$
implying
$$\ln|\cos\pi\phi_n|=-\ln 2-\sum_{m=1}^\infty {(-1)^m\over m}\cos
(2\pi\phi_nm).\eqno(2.4)$$
The original sum (2.1) reduces to the form
$$A_N=-\sum_{m=1}^\infty {(-1)^m\over m}\sum_{n=0}^{N-1}\cos
(2\pi\phi_nm).\eqno(2.5)$$
If the phases $\phi_n$ are random
$$\langle A_N\rangle =0\eqno(2.6)$$
while
$$\langle (A_N)^2\rangle ={N\over 2}\sum_{m=1}^\infty {1\over
m^2}={\pi^2\over 12}N\eqno(2.7)$$
where $\langle\ \rangle$ denote averages over the realizations of the
random potential.

Therefore, if the phase $\phi_n$ is random one expects that the typical
size of $A_N$ will be of the order $\sqrt N$. \ If, on the other
hand, $\phi_N$ is given by (2.1) the behavior is completely different.  In
this case, the sum over $n$ is just a geometric series, namely
$$\sum_{n=0}^{N-1}\cos 2\pi\alpha nm={1\over 2}\left[
{\sin\pi\alpha (2N-1)m\over \sin\pi\alpha m}+1\right]\eqno(2.8)$$
leading to
$$A_N=\tilde A_N+{1\over 2}\ln 2\eqno(2.9)$$
with
$$\tilde A_N=-{1\over 2}\sum_{m=1}^\infty {(-1)^m\over m}{\sin\pi\alpha(2N-
1)m\over \sin\pi\alpha m}\eqno(2.10)$$
The sum $\tilde A_N$ is dominated by the terms where the denominators are
small.  These are terms where $m$ is equal to $q_k-$ a denominator of the
rational approximant of order $k$ to $\alpha$. \ The method that will be
used is closely related to the one that was applied by Berry$^{18}$
for a different problem.  Let $\alpha$ be an irrational number,
and the elements of its continued fraction expansion be $a_0,a_1\dots
a_i\dots$ and let its rational approximants of order $k$ be
$$\alpha_k=p_k/q_k\eqno(2.11a)$$
where $p_k$ and $q_k$ are integers.  The error in this approximant is
$$\epsilon_k=\alpha-\alpha_k. \eqno(2.11b)$$
For almost all irrational numbers (Ref.~ 19, p.~78; Ref.~20, p.~169)
$$|\epsilon_k|\sim {\overline c_2\over q_k^2\ln q_k}\eqno(2.12)$$
while for irrational numbers with continued fractions with bounded elements
$a_i$, (Ref.~ 19, p.~44)
$$|\epsilon_k|\sim {c_2\over q_k^2}\eqno(2.13)$$
where $\overline c_2$ and $c_2$ are constants independent of $k$. \ For the
gold
en
mean $\alpha={\sqrt{5}-1\over 2}$ for example $c_2=1/\sqrt 5$ (Ref.~19, p.~41;
Ref.~20, p.~163).

The sum (2.10) can be estimated as
$$\tilde A_n=-{1\over 2}\sum_k{(-1)^{q_k}\over
q_k}{\sin \pi\epsilon_k(2N-1)\over \sin\pi\epsilon_k}.\eqno(2.14)$$

For irrational numbers with continued fractions with bounded elements the
approximation (2.13) holds resulting in

$$\tilde A_N=-{1\over 2}\sum_k{(-1)^{q_k}\over
c_2\pi}\cdot\sin\left[{\pi(2N-1)\over q_k}c_2\right]\eqno(2.15)$$

We turn now to estimate the sum $\tilde A_N$ of (2.15) for large $N$.
The contributions from the terms with large values of $q_k$ can be
approximated by a convergent integral.  This motivates to split the sum
in two parts, namely,
$$\tilde A_N=-{1\over 2c_2\pi}[\tilde A_N^{(1)} +\tilde A_N^{(2)}]\eqno(2.16)$$
where
$$\tilde A_N^{(1)}=\sum_{k=0}^{k_N}(-1)^{q_k}\sin\left[{\pi(2N-1)\over q_k}
c_2\right]\eqno(2.17)$$
and
$$\tilde A_N^{(2)}=\sum_{k=k_N+1}^{\infty}(-1)^{q_k}\sin\left[
{\pi(2N-1)\over q_k}
c_2\right]\eqno(2.18)$$
The condition
$$q_{k_N}=N\eqno(2.19)$$
determines $k_N$.  We first turn to evaluate the sum $\tilde A_N^{(2)}$.

The factor $(-1)^{q_k}$ introduces a rapid oscillation in the
terms of the sum.
However locally the number of even values of $q_k$ is
not equal to the number of odd values.
Actually, the number of odd values of $q_k$ is twice
the number of even values as demonstrated in Appendix~A. \
Therefore this factor does not result
in local cancellations. It may change the value of the sum but
not its asymptotic $N$ dependence.  We will ignore this factor in the
evaluation of $\tilde A_N^{(2)}$. \
If $(-1)^{q_k}$ is ignored the sum can be approximated by an
integral.  For this purpose we note that
$$q_k\sim e^{\delta k}\eqno(2.20)$$
where $\delta$ depends in general on $\alpha$. \ For the golden mean
$$\delta=-\ln\alpha\eqno(2.21)$$
while for generic numbers (almost all irrational numbers for which Eq.(2.12)
applies),
$$\delta={\pi^2\over 12\ln2}.\eqno(2.22)$$
For the numbers with bounded $a_i$ like the golden mean, the
sum behaves as
$$\eqalign{\tilde A_N^{(2)}&\sim \int_{k_N}^\infty dk\sin (2\pi
Nc_2/q_k)\cr
&\sim {1\over \delta}\int_{\overline q}^\infty {dq\over q}\sin(2\pi N/q)\cr}
\eqno(2.23)$$
where $\overline q=q_{k_N}/c_2$  and $k_N$ are related via (2.21) and
(2.19). \
This integral is convergent. \ It can
be easily estimated with the help of the change of variable
$$x={1\over q}\eqno (2.24)$$
leading to
$$\tilde A_N^{(2)}\sim{1\over\delta}\int_{0}^{1/\overline q} dx
{\sin 2\pi Nx\over
x}\eqno(2.25)$$
Since
$$\lim_{N\to\infty}{\sin 2\pi Nx\over x}=\pi\delta(x)\eqno(2.26)$$
$\tilde A_N^{(2)}$ behaves as a constant in the limit $N\to\infty.$

The error in the estimate of the sum $\tilde A_N^{(2)}$ by an integral is of
the
order of
$$E_N=\sum_{k=k_N+1}^\infty {c_2\pi(2N-1)\over q_k}={c_2\pi(2N-1)e^{-\delta(k_N
+1)}\over 1-e^{-\delta}}\eqno(2.27)$$
that is bounded following the definition (2.19) of $k_N$ and (2.20).  Therefore
the sum $\tilde A_N^{(2)}$ is bounded with a bound that is finite in the limit
$N\to\infty$. \ We turn now to estimate the behavior of $\tilde A_N^{(1)}$,
that turns out to dominate the sum $\tilde A_N$ of (2.15).  For this
purpose it is rewritten in the form:
$$\tilde A_N^{(1)}=\sum_{m=0}^{k_N}(-1)^{q_k}\sin 2\pi\mu\gamma^m\eqno(2.28)$$
where
$$2\mu=c_2(2N-1)e^{-\delta k_N}\eqno(2.29)$$
$$\gamma=e^\delta\eqno(2.30)$$
and
$$k=k_N-m\eqno(2.31)$$
Note that $\mu$ approaches a constant in the limit $N\to\infty$.

Typically $\mu$ is an irrational number.  For integer values of $\gamma$, the
sequence
$$\phi^\mu(m)=(\mu\gamma^m){\ \rm mod\ } 1\eqno(2.32)$$
can be considered random, since it is a shift along the digits of an
irrational number, in the base $\gamma$.  If $\gamma$ is not an integer it is
hard to believe that $\phi^\mu(m)$ will be less random (although we do not
claim
to justify it rigorously).  In what follows it will be assumed that
$\left\{\phi^m(m)\right\}$ is a random uncorrelated sequence.  With this
assumption $\tilde A_N^{(1)}$ takes the form
$$\tilde A_N^{(1)}=\sum_{m=0}^{k_N}(-1)^{q_k}\sin 2\pi\phi^\mu(m)\eqno(2.33)$$

Assuming that $\left\{\phi^\mu(m)\right\}$ is random one finds that the
variance of $\tilde A_N^{(1)}$ is ${1\over2}k_N\sim\ln N/2\delta$. Hence
the typical magnitude of $\tilde A_N^{(1)}$ is of the order of $\sqrt{
\ln N}$. \ Therefore $\tilde A_N^{(1)}$ is the dominant contribution to
$\tilde A_N$ of (2.15) and $\tilde A_N^{(2)}$ is negligible for large $N$.
Because of the randomness of $\left\{\phi^\mu(m)\right\}$ the sums
$\tilde A_N^{(1)}$ and $\tilde A_N$ look as a random walk is one dimension.

It is instructive to define $\tilde A_N^{max}$ as the maximal value that
$|\tilde A_{N^\prime}|$ may take for $N^\prime\le N.$ \ From the random walk
property of $\tilde A_N^{(1)}$ one finds that for large $N$
$$\tilde A_N^{max}\sim k_N\sim \ln N,\eqno (2.34)$$
[It is easy to show that the typical interval between two consecutive
$N^\prime$'s for which \break
$\tilde A_{N^\prime}^{(1)}> C\ln N^\prime\ (C<\delta)$ is
$\Delta N\le N^{\prime 2C^{2}\delta}(\pi\ln N^\prime/\delta)^{1/2}<<
N$].

For generic $\alpha$, where (2.12) holds the sum (2.15) should be replaced
by,
$$\tilde A_N=-{1\over2}\sum_k{(-1)^{q_k} \ln q_k\over
\overline c_2\pi}\sin\left[{\pi (2N-1)\overline c_2\over q_k\ln
q_k}\right]\eqno(2.35)$$

The sum can be split in two parts as was done in (2.16) and sums similar
to (2.17) and (2.18) are obtained with each term multiplied by
$\ln q_k$ and $c_2$ replaced by $\overline c_2$.  The sum $\tilde A_N^{(2)}$
is approximated by an integral with an error of the order (2.27) that is
bounded .

Considerations similar to those leading to (2.25), with the change of
variable
$$x={\overline c_2\over q\ln q}\eqno(2.36)$$
lead for large $\overline q$ to
$$\tilde A_N^{(2)}\sim-{1\over\delta} \int_0^{\overline c_2/\overline q\ln
\overline q} dx {\ln x\over x}\sin 2\pi
Nx\eqno(2.37)$$

Here $\overline q=q_{k_N}=N$. \ Since for large $N$ the upper limit of the
integral is much smaller that $1/N$ the integral can be estimated as
$$\tilde A_N^{(2)}\sim-{2\pi N\over\delta}\int_0^{\overline c_2/\overline
q\ln\overline q}dx\,\ln x\sim -{2\pi\over \delta}{N\over\overline q}=
{2\pi\over \delta}\eqno(2.38)$$
that is bounded.  Again the leading contribution results from $\tilde
A_N^{(1)}$
that now takes the form
$$\tilde
A_N^{(1)}=\delta\sum_{m=0}^{k_N}(-1)^{q_k}k\sin\left[{2\pi\mu\gamma^m\o
ver
\delta k}\right]\eqno(2.39)$$
where $\mu$, $\gamma$, and $k$ are defined by (2.29 -- 2.31).  Following the
argument presented after Eq.~(2.32) the sequence $\tilde A_N^{(1)}$ can be
considered random with the variance of the order of ${1\over 2}(\delta k_N)^2
\sim {1\over 2}(\ln N)^2$. \ The typical values of this sequence are therefore
of the order of $\ln N$, while the maximal value of $|\tilde A_{N^\prime}|$
for $N^\prime\le N$ is,
$$\tilde A_N^{max}\sim k_N^2\sim (\ln N)^2\eqno(2.40)$$

We conclude that the typical growth of $A_N$ with $N$ is $\ln N$ and not
$\sqrt N$ (as if the original phases $\phi_n$ were random).  For irrational
numb
ers with
continued fractions with bounded elements the typical growth is of the order
$\sqrt{\ln N}$. \ The maximal value that the sum takes, grows as $(\ln N)^2$
and $\ln N$ respectively in these cases.

In Fig.~1 $A_N$ and $A_N^{max}$ as a function of $N$ are depicted for $\alpha =
{1\over 2}(\sqrt 5 - 1)$ and $\alpha = {1\over e}$. \ The values of $A_N$ are
of
 the order unity and are very
different from values expected for random sequencing.  It is hard to see the
expected growth of $A_N$ because of its erratic nature.  Since $\tilde
A_N^{max}
$
is a monotonically increasing sequence, its growth is systematic and obvious.
Note the difference between the $\ln N$ and a faster growth, consistent with
$(\ln N)^2$, found for irrationals with bounded and unbounded elements of the
co
ntinued
fractions.  These results are in accord with the assumptions that were made
on the phases of the sine in (2.33) and (2.39).

In Fig. 1(b) we see that in addition to the continuous increase of
$A_N^{\rm max}$ with $\ln N$, there are also discontinuous ``jumps"
periodically
spaced and of similar heights.  These jumps are compounded by contributions
from two consecutive N's.  When we looked instead to the series $B_N=\ln\pi^N_
{r=1}|\sin(\pi\alpha r)|$ with the same $\alpha = (\sqrt 5 - 1)/2$ we observe
a continuous increase with $\ln N$ (but with a different slope which may be
understood from the absence of the factors $(-1)^m$ in Eq. (2.10)), but without
any discontinuous jumps.  For other rational $\alpha$'s other structures of
periodic jumps were seen.  On the other hand no periodic structure is seen in
these jumps for non-algebraic $\alpha$'s (Fig. 1(d)).  This is clearly a
commens
urability
effect connected with the continuous fraction expansion of $\alpha$.  So far
we do not have a more quantitative understanding of these discontinuities.

All computations were done in double precision.  To check possible effects of
the finite precision we ran the same program also in single precision (and
with both degrees of precision on a different machine) without discernible
differences.  So, we are confident that the structure observed is real and is
not an artifact of the finite precision.

In this section we examined so far the behavior of $A_N$ for $\nu=1$
namely the expression (2.2) for the phase $\phi_n^\nu$ was used.
In the next chapter values of $\nu\not=1$ will be
investigated.  As a first step in that direction we explore the stability
of the $\nu=1$ behavior for slight deviations of $\nu$ from this value.

We therefore consider the case $\nu=1+\epsilon$ and attempt to find the
deviation:
$$\Delta A_N^\epsilon(\alpha)=A_N^{1+\epsilon}(\alpha)-
A_N^1(\alpha).\eqno(2.41)$$
A divergence in the $\Delta A_N^\epsilon(\alpha)$ as $N\to\infty$ for
arbitrarily small $\epsilon$ will indicate the instability of the $\nu=1$
behavior.

For small $\epsilon$ we expand:
$$\eqalign{A_N^{1+\epsilon}(\alpha)&=\sum_{r=1}^N \ln|\cos(\pi\alpha
r^{1+\epsilon})|\cr
&=\sum_{r=1}^N \ln|\cos\{\pi\alpha r(1+\epsilon\ln r\})|\cr
&=\sum_r\ln|\cos(\pi\alpha r)|+\epsilon\pi\alpha r\ln
r|tg(\pi\alpha r)|\cr}\eqno(2.42)$$
Thus from (2.26) we have:
$$\Delta A_N^\epsilon(\alpha)=\epsilon\pi\alpha\sum_{r=1}^N r\ln r
|tg(\pi\alpha r)|.$$

It is already clear that this sum will diverge at least as $N^2\ln N$. \
However, the presence of the $tg (\pi\alpha r)$ may yield a stronger
divergence.

To explore this effect we concentrate on the points $r$ such that $2\alpha
r\approx 2n+1$. \ Near these points $2\alpha$ is approximated by the
continued fraction ${p_k\over q_k}$ so that $|2\alpha-{p_k\over q_k}|\sim
\delta_k$, with $\delta_k\sim {1\over q_k^2\ln q_k}$. \ Evidently the
singular points will be these $r=q_k$ for which $p_k$ is odd.  Near these
points
$$tg\left[(n+{1\over 2})\pi +{\pi\delta_k\over 2}\right]\sim
{2\over \pi\delta_k}\eqno(2.43)$$

Inserting to $\Delta A_N^\epsilon(\alpha)$ we obtain,
$$\eqalign{\Delta A_N^\epsilon(\alpha)&\sim {\sum_k}^\prime q_k\ln
q_k {2q_k^2\ln q_k\over\pi q_k}\cr
&\sim {\sum_k}^\prime (q_k\ln q_k)^2\cr}\eqno(2.44)$$
where $\sum^\prime$ means sum only over these $k$ for which $p_k$ is odd.

The asymptotic behavior is obtained by transforming to an integral as
before:
$$\Delta A_N^\epsilon(\alpha)\sim \epsilon N^2(\ln N)^2\eqno(2.45)$$
For the non-generic rational numbers like the Golden mean there will be one
less power of $\ln N$ hence $\Delta A_N^\epsilon (G.M.)\sim N^2\ln N$.
\ In any case, the $\nu=1$ point is unstable to any small deviation in
$\nu$ since the limits $N\to\infty$ and $\epsilon\to0$ do no commute.

In the following section the sums $A_N^\nu$ will be investigated for
various values of $\nu$.
\vskip 10pt

\noindent {\bf III. \ The Sums $A_N^\nu$ for Various Values of $\nu\not=1$}
\vskip 10pt

In this section the sums
$$A_N^\nu=\sum_{n=0}^{N-1}\ln 2|\cos\pi\phi_n^\nu|\eqno(3.1)$$
with
$$\phi_n^\nu=\alpha n^\nu\eqno(3.2)$$
will be investigated.

With the help of (2.4) that is independent of the form of $\phi_n^\nu$ it can
be written in the form
$$A_N^\nu=-\sum _{m=1}^\infty {(-1)^m\over m}S_N^\nu(\alpha m)\eqno(3.3)$$
where
$$S_N^\nu(\alpha)=\sum_{n=0}^{N-1}\cos 2\pi \alpha n^\nu\eqno(3.4)$$
The sum $A_N^\nu$ is dominated by the $S_N^\nu(\alpha m)$ with the lowest
values of $m$.  The general form of these sums does not depend strongly
on $m$. \ Sums of the form (3.4) were encountered in earlier work.$^{8}$ \
{}From
the experience of these investigations it is known that for $\nu\not=1$ there
are the following distinct regimes$^{8-10, 13-16}$ {\bf A.} \ $0<\nu<1$, {\bf
B.} \ $1<\nu<2$,
{\bf C.} \ $\nu>2$ and {\bf D.} \ $\nu=2$.
We will investigate the sums in these regimes separately.  It turns out
that the regime $0<\nu<1$ should be divided in two subregimes and the behavior
of the sum (3.3) for $0<\nu<{1\over2}$ differs from the one found for
${1\over2}
<\nu<1$.

\vskip 10pt

\noindent {\it A. \ The Regime $\nu<1$}
\vskip 10pt

For $\nu<1$ the terms in the sums $S_N^\nu(\alpha)$ of (3.4) vary slowly
with $n$ for sufficiently large $n$. \ Since each term in the
sum is of order unity the sum $S_N^\nu(m\alpha)$ can be
approximated by an integral in the regime where the difference between
two consecutive terms is less than unity.  This requires
$$2\pi\alpha m\nu<<n^{1-\nu}.\eqno(3.5)$$
The terms in the sum (3.3) can be considered as points $(n,m)$ on a
two dimensional lattice, with the restrictions $0<n<N-1$ and $1<m<\infty$. \
We define a line in the $(n,m)$ plane
$$2\pi\alpha m\nu=c_3n^{1-\nu}\eqno(3.6)$$
so that on one side of this line (3.5) is satisfied and the sums (3.4) can
be replaced by integrals while on the other side this is impossible and
a different approximation is required.  $c_1$ in (3.6) is of order unity.
The sum of (3.3) can be divided into two parts so that in each part all terms
are on one side of (3.6), namely
$$A_N^\nu=-B_n^\nu-D_N^\nu\eqno(3.7)$$
where
$$B_N^\nu=\sum_{m=1}^M {(-1)^m\over m} \sum_{n=n^*(m)}^{N-1}\cos 2\pi
\alpha m n^\nu\eqno(3.8)$$
and
$$D_N^\nu=\sum_{n=0}^{N-1}\sum_{m=m^*(n)}^\infty {(-1)^m\over m}\cos 2\pi
\alpha mn^\nu.\eqno(3.9)$$
The limits of the summations are
$$m^*(m)=\Int [c_1 n^{1-\nu}/2\pi\alpha\nu]+1\eqno(3.10)$$
$$n^*(m)=\Int [(2\pi\alpha m\nu/c_1)^{1/(1-\nu)}]+1\eqno(3.11)$$
and
$$M=\Int[c_1(N-1)^{1-\nu}/2\pi\alpha\nu]\eqno(3.12)$$
where $\Int[x]$ denotes the integer part of $x$ namely the largest integer
that is smaller than $x$.

The sum over $n$ in (3.8) can be approximated by an integral namely
$$\Delta S_{n^*, N}(\alpha m)=S_N^\nu(\alpha m)-S_{n^*}^\nu(\alpha m)
=\sum_{n=n^*(m)}^{N-1}\cos 2\pi\alpha mn^\nu\approx\int_{n^*}^{N-1}
dn\,\cos 2\pi m\alpha n^\nu\eqno(3.13)$$
Introducing the variable $x=n^\nu$ one obtains
$$\Delta S_{n^*,N}^\nu (\alpha m)\approx {1\over \nu}\int_{x^*}^X dx\,
x^{{1\over\nu}-1}\cos 2\pi\alpha mx\eqno(3.14)$$
where
$$X=(N-1)^\nu\eqno(3.15)$$
and
$$x^*=n^{*\nu}\eqno(3.16)$$
Integrating by parts one finds for large $N$
$$\eqalign{\Delta S_{n^*,N}^\nu
(\alpha m)&\sim {X^{{1\over \nu}-1}\over 2\pi\alpha m\nu}
\sin 2\pi\alpha m X-{x^{*{1\over \nu}-1}\over 2\pi\alpha m\nu}
\sin 2\pi\alpha m x^*+R_{N,m}\cr
&={N^{1-\nu}\over 2\pi\alpha m\nu}\sin 2\pi\alpha m N^\nu-
{n^{*1-\nu}\over 2\pi\alpha m \nu}\sin 2\pi\alpha m n^{*\nu}+R_{N,m}\cr}
\eqno(3.17)$$
Where the remainder term is

$$R_{N,m}={\cal O}\left({N^{1-2\nu}\over m^2}\right)+{\cal O}
\left({n^{*(1-2\nu)}\over
m^2}\right)\eqno(3.18)$$

The sums $\Delta S_{n^*,N}(\alpha m)$ should be substituted in (3.8)
in order to calculate $B_N^\nu$. \ We will show first that the contribution
of the second term in (3.17) is negligible for large $N$. \ Its contribution
to $B_N^\nu$ is
$$\eqalign{R_B&=-\sum_{m=1}^M {(-1)^m\over m^2}{(n^*)^{1-\nu}\over 2\pi
\alpha\nu}\sin(2\pi\alpha m n^{*\nu})\cr
&\approx -{1\over c_1}\sum_{m=1}^M {(-1)^m\over m}\sin(2\pi\alpha mn^{*\nu})\cr
}\eqno(3.19)$$
where (3.11) was used.

This sum is clearly bounded by a term that grows as $\ln M$ that is
proportional to $\ln N$ for large $N$ due to (3.12). \ The estimate
of $R_B$ can be improved since for $0<\nu<1$, $1<{1\over 1-\nu}<\infty$
and the phase of the sine in (3.19) varies rapidly with $m$ (see (3.11)). \
The contribution of a group of terms in the vicinity of some $m$ is of the
order of $1/m^2$, consequently the sum is convergent in the limit
$M\to \infty$ and therefore it behaves as a constant for large $N$. \ The
contribution of the remainder term $R_{N,m}$ of (3.18) is of the order of
$N^{1-2\nu}$ since the sum $\sum_{m=1}^\infty {(-1)^m\over m^2}$ is
convergent.  Therefore, substitution of (3.17) in (3.8) yields
$$B_N^\nu={N^{1-\nu}\over 2\pi\alpha\nu}\sum_{m=1}^M{(-1)^m\over m^2}\sin
2\pi\alpha mN^\nu+R_N\eqno(3.20)$$
where $R_N$ is a remainder term of the order $N^{1-2\nu}$. \ The sum over
$m$ is convergent (in the limit $M\to\infty$), therefore it is dominated
by the small $m$ terms.  The remainder term $R_N$ is negligible for
sufficiently large $N$.

We turn now to analyze the behavior of $D_N^\nu$ of (3.9). \ The terms in the
sum vary rapidly with $m$. \ If one groups several terms in the vicinity of
$m$, their sum is of the order $1/m^2$. \ Consequently the sum over $m$
in (3.9) is approximately
$$\tilde S_n^\nu=\sum_{m=m^*(n)}^\infty {(-1)^m\over m}\cos 2\pi
\alpha mn^\nu\sim {\eta_1(n)\over m^*(n)}\eqno(3.21)$$
where $\eta_1(n)$ is a function of $n$ with a bounded amplitude.  It
is determined mainly by the terms in the vicinity
of $m^*$ and the function $\Int[x]$ of (3.10). \ In the derivation
we used the fact that up to a constant $\sum_{m=m^*}^\infty
{1\over m^2}\sim {1\over m^*}$. \ With the help of (3.10) one finds that
for large $N$ the sum (3.9) is approximately
$$D_N^\nu\sim{2\pi\alpha\nu\over c_1}\sum_{n=0}^N{\eta_1(n)\over n^{1-\nu}}\sim
{2\pi\alpha\nu\over c_1}\int_{\overline n}^N dn{\eta_1(n)\over n^{1-\nu}}
\eqno(3.22)$$
where $\overline n$ is some lower cutoff that is not important since the
integral diverges in the limit $N\to\infty$ and is therefore dominated by the
vicinity of $N$. \ The function $\eta_1(x)$ is determined mainly by the terms
in the vicinity of $m^*$ in the sum (3.21), where $m^*$ is determined
by (3.10). \ One
can see from direct examination of these, that the values of $n$ where function
$\eta_1(n)$ changes sign became sparser as $n$ increases.
This function is expected to exhibit random like behaviors on very
long scales due to the nature of the functions $\Int[x]$ and $\cos x$ and the
slow variation of their argument with $n$. \ Because of the slowness of the
variation of $\eta_1(n)$ for large $N$, the asymptotic behavior of the
integral (3.22) is
$$D_N^\nu\sim {2\pi\alpha\nu\over c_1}\eta_1(N)\int_{\overline n_1}^N
{dn\over n^{1-\nu}}\sim \eta(N) N^\nu\eqno(3.23)$$
where $|\eta(N)|\approx{2\pi\alpha\over c_1}$ and $\overline n_1$ is some lower
cutoff that has no significance.

The asymptotic form of the sum (3.3) is
$$A_N^\nu\sim -{N^{1-\nu}\over 2\pi\alpha\nu}\sum_{m=1}^M {(-1)^m\over m^2}
\sin 2\pi\alpha m N^\nu-\eta(N)N^\nu\eqno(3.24)$$
as one finds from substitution of (3.20) and (3.23) into (3.7). \ The first
term is an oscillating function of $X=N^\nu$ with the period
$1/\alpha$ while the second is a slowly varying ``randomlike" term.  The ratio
o
f
their magnitudes is of the order
$$N^{1-2\nu}\lim_{N\to\infty}\cases{\infty&for
$\nu<1/2$\cr 0&for $\nu>1/2$.\cr}\eqno(3.25)$$
Therefore the first term dominates for $\nu<1/2$ while the second dominates for
$\nu>1/2$. \ In order to investigate how well the approximation (3.24)
works we plotted in Fig.~2, $A_N^\nu$ of (3.1) for several values of $\nu$. \
Actually, we plotted
$$F_\nu(x)={A_N^\nu\over N^{1-\nu}}\eqno(3.26)$$
as a function of $X=N^\nu$. \ In Fig.~3 the Fourier transform $\hat F_\nu(Q)$
of $F_\nu(x)$ is presented.  It is clear that the period is indeed
$1/\alpha$. \ For $\nu\le1/2$ the sum is purely periodic in $X$ and the
weight of the high harmonics, corresponding to large values of $m$ in the sum
(3.24) is small.  For $\nu>1/2$ the erratic behavior
resulting from the nature of the function $\eta(N)$ is obvious.
Since it is slow it results in a peak at $Q=0$ of the Fourier transform.
The oscillating behavior resulting from the first term is observed as well.
For $\nu=1$ the periodic structure that is found for $\nu<1$ disappears as is
cl
ear
from Fig.~4.
\vskip 10pt

\noindent {B. \ The regime $1<\nu<2$}
\vskip 10pt

In this regime the sums are dominated by terms in the vicinity of $n$ that
satisfy the condition$^{8,9}$
$$2\pi\alpha\nu n^{\nu-1}\approx 2 l\pi\eqno(3.27)$$
where $l$ is an arbitrary integer.  For large $n$ the distance between
two consecutive values of $n$ that satisfy (3.27) is
$$\Delta n=n^{2-\nu}/\alpha\nu(\nu-1)\eqno(3.28)$$
Note that $\Delta n\to\infty$ in the limit $n\to \infty$. \ For large
$N$ the typical size of $S_N^\nu(\alpha)$ in this regime is
$\sqrt{N}$ as if $\phi_n^\nu$ would be random as demonstrated in Fig.~5. \
Note that the number of terms in the sums of this figure
is of the order of $10^4$. \
Although the typical size is similar to the one found for a random
phase, the variations differ strongly from those found for random phases.
The variations are large and take place for values of $n$ that satisfy
(3.27) and the resulting separation between them is (3.28). \ The
growth takes place in extremely narrow regions and is most pronounced for
small $m$ where the separation between these regions is the largest.  (For
$m\not=1$, $\alpha$ in (3.28) should be replaced by $m\alpha$.)  The
sum (3.3) is dominated by terms with small $m$, therefore the separation
between regions of the largest variation is given approximately by (3.28).
This is demonstrated in Fig.~5, where the variation in a relatively
small region is plotted. \ The prediction of (3.28) for separation between
the regions of large variation in Fig.~5 is $\Delta n=
2.15\times 10^3$ in good agreement
with the numerical results.
\vskip 10pt

\noindent{\it C. \ The regime $\nu>2$}
\vskip 10pt

In this regime the sums $S_N^\nu(\alpha)$ are expected to behave as if the
phases $\phi_n^\nu$ are random.$^{8, 13-16}$ \
Moreover, in this regime ($\phi_n^\nu)
{\rm mod1}$ is found to pass the $\chi^2$ test for psuedorandomness.$^{16}$ \
It is expected that $A_N^\nu$ of (3.1) will behave as if the phases are
random.  This is demonstrated in Fig.~6 where the behavior of $A_N^\nu$ with
$\nu=3$ is depicted.  It is consistent with a $\sqrt N$ increase.  This plot
could not be distinguished from that obtained with ``random" phases.  (We
recall
,
however, that the random number generator used to generate the random phases
is also based on a quasiperiodic sequence).
\vskip 10pt

\noindent{\it D. \ $\nu=2$}
\vskip 10pt

This is a bordering case between regimes B and C.  For $\nu=2$,
($\phi_n^\nu$)mod1 fails the $\chi^2$ test for
psuedorandomness but not remarkably and as tests of pseudorandomness are
concerned it is quite close to a random sequence.$^{16}$ \ In
particular $S_N^\nu$
exhibits fluctuations consistent with $\sqrt{N}$ but there are no regions of
strong variation of the form that were found for $1<\nu<2$. \ The sum $A_N^\nu$
is depicted in Fig.~6b. (note its difference from that with $\nu = 1.5$
plotted in Fig.~5).
\vskip 10pt

\noindent{\bf 4. \ Conclusions}
\vskip 10pt

To summarize, our findings on $A_\nu(t)=\ln |J_\nu(t)|$ are:

\item{A.} For $\nu=1$:  The behavior depends on the number theoretic properties
of the ratio $\alpha=\phi/\phi_0$. \  For algebraic irrational $\alpha$, with
bounded continued-fraction expansion, $|J^\nu(t)|$ increases algebraically
as $t^\beta$.
For a generic rational number, such that its continued fraction expansion
consists of an unbound series of integers which are as likely to be even or
odd, $|J^\nu(t)|$ increases faster than algebraically with analytical
prediction of an increase like $t^{\beta \ln t}$. \ It remains to
understand the dependence of the exponents $\beta(\alpha)$ on $\alpha$,
which we leave for future studies.  The same holds for the discontinuous jumps
discussed at the end of Ch.~2.

\item{B.}For $\nu\not=1$ different regimes of behaviors were identified:
\itemitem{{\it i.}}$0<\nu<{1\over 2}$:  The behavior of $A_N^\nu$ is
essentially
periodic in the variable $X=t^\nu$
with a finite number of harmonics dominates the whole sum.
\itemitem{{\it ii.}}${1\over2}<\nu<1$:  In this regime there appears a large
scale random component superimposed on the periodic behavior as exhibited in
the Fourier spectrum by noise band near the origin.
\itemitem{{\it iii.}}For $1<\nu<2$:  The behavior of the series is dominated by
the special points where $\alpha\nu n^{\nu-1}$ is approximately integer.
\itemitem{{\it iv.}}For $\nu>2$ the behavior of $J^\nu(t)$ is indistinguishable
from that of a product of cosines with a random phase uniformly distributed
in $[0,2\pi]$.

As for possible experimental realizations:  for $\nu=1$ the best candidates are
artificially fabricated superconducting (or Josephson junctions)
networks.$^{21-23}$  The
existence of surface states (localized solutions of the linearized
Ginzburg-Landau equations) was already discussed in the context of numerical
simulations of these networks.$^{24}$ \
Measurements of the exponential decay of the
supercurrent transmission as function of the magnetic field and/or
the sample width, may exhibit some
of the properties analyzed here.

It will be interesting to go beyond the ``directed path'' approximation
and see how inclusion of returning loops modifies
this behavior.  Preliminary results indicate that the important features
will be conserved as long as the energy is in the gap (and, hence, returning
loops are finite). \

Finally crossing the ``mobility edge" between localized and extended states
would allow to make the connection between our results and the previous
studies of the Hofstadter bulk eigenstates.
\vskip 10pt

\noindent{\bf Acknowledgements}
\vskip 10pt

It is our pleasure to thank M. Kardar and E. Medina for useful discussions.
We are also grateful to the anonymous referee whose comments helped us
clarify some issues and avoid trivial mistakes. This work was supported in
part by the US-Israel Binational Science Foundation (BSF).  YS acknowledges
the support of the Compton Foundation during his visits to the Technion and
the hospitality of Ora Entin-Wohlman during his visit to Tel-Aviv University.
Support from Minnesota University Supercomputer Institute is also acknowledged
(XRW).

\vfill\eject

\centerline{Appendix A:  Parity Statistics of the Approximants' Denominators
$q_k$}
\vskip 10pt

In this appendix we show that for a generic irrational number, such that
the parity of the integers in its continued-fraction expansion are randomly
even or odd without correlations,  the $q_k$'s are twice as likely to be
odd than even in the $k\to\infty$ limit.

To use this in Eq.~(2.15) we also need to show that $p_k$ and $q_k$ are never
both even (in which case they could be both divided by two thus distorting
the parity ``statistics").

We recall that the recursion relations for $q_k$ and $p_k$ in terms of the
contined-fraction integers $a_k$ are
$$q_k=a_kq_{k-1}+q_{k-2}\qquad {\rm with} \qquad q_0=1,\ q_1=a_1\eqno(A1)$$
$$p_k=a_kp_{k-1}+p_{k-2}\qquad {\rm with} \qquad p_0=a_0,\ p_1=a_0a_1+1\eqno
(A2)$$

We begin by considering the parities $\pi_k=\pi(q_k)$, $\pi_k=0(1)$ if
$q_k$ is even (odd).

Let us define the two components vectors $\vec v^{(k)}=\pmatrix{\pi_{k-1}\cr
\pi_{k-2}\cr}$ over the $Z_2$ field $\{0,1\}$.
The whole vector field consists of four elements
$$\vec e_1=\pmatrix{1\cr 1\cr}, \vec e_2=\pmatrix{0\cr 1\cr},
\vec e_3=\pmatrix{1\cr 0\cr}, \ {\rm and}\  \vec e_4=\pmatrix{0\cr 0\cr}.$$
The recursion relation (A1) for $q_k$ defines a linear map between
$v^{(k)}$ and $v^{(k+1)}$ which only depends on $\pi(a_k)$. \
If we denote by $\omega_i^{(k)}$ the probability for $v^{(k)}=\vec e_i$,
$i=1,2,3,4$; the probability distribution for $\pi(a_k)$ will thus
determine a relation between $\omega_i^{(k+1)}$ and $\omega_i^{(k)}$. \

For a generic irrational number, the probability for $a_k$ to be equal to m is
$1/m(m-1)$.  Hence the probability for $a_k$ to be odd is $\ln 2$.
\ The matrix relating $\omega_i^{(k)}$ for consecutive $k$'s is found to be:

$$\pmatrix{{1-ln 2}&0&{ln 2}&0\cr
{ln 2}&0&{1-ln 2}&0\cr
0&1&0&0&\cr
0&0&0&1\cr}\eqno(A3)$$

Obviously $\omega_4$ = 0 because $p_k$ and $q_k$ cannot have a common
divisor$^{19}$.

To find the stationary distribution for $\omega_i^{(\infty)}$ \ $i=1,2,3$
we note that A3 defines a Markov process and the eigenvector with the largest
eigenvalue (by virtue of the Perron-Frobeni\"us theorem) has
$\omega_1=\omega_2=\omega_3$ which (since $\omega_4=0$) is
$\vec\omega^{(\infty)}=(1/3, 1/3, 1/3, 0)$. \ To assume that this
distribution is reached we still need to assure that $\vec \omega^{({\rm in})}$
(the initial $\vec\omega$) is
not orthogonal to $\vec\omega^{(\infty)}$. \ Indeed $\vec
v^{({\rm in})}=\pmatrix{1\cr \pi(a_1)\cr}$ and therefore
$\vec\omega^{({\rm in})}=(1/2,
0, 1/2, 0)$ and $\vec\omega^{({\rm in})}\cdot \vec\omega^{(\infty)}=1/3>0.$ \

Counting the numbers of times 1 and 0 appears in $\vec e_1$, $\vec e_2$,
and $\vec e_3$ we conclude that if they are uniformly distributed as
$k\to\infty$, the probability for $q_k$ to be odd is twice as large that to
be even.

\vfill\eject

\noindent{\bf References}
\vskip 10pt

\item{1.}D. R. Hofstadter, Phys. Rev. {\bf B14}, 2239 (1976).
\item{2.}G. H. Wannier, Phys. State. Solidi, {\bf B88}, 757 (1978).
\item{3.}M. Ya. Azbel, Zh. Eksp. Theor. Fiz. {\bf 48}, 929 (1964).
\item{4.}For a recent review see Y. Shapir and X. R. Wang, Mod. Phys.
Letts. B{\bf 4}, 1301 (1990).
\item{5.}E. Medina, M. Kardar, Y. Shapir and X. R. Wang,
Phys. Rev. Letts. {\bf 84}, 1916 (1990).
\item{6.}Y. Shapir, X. R. Wang, E. Medina and M. Kardar, in {\it Hopping
and Related Phenomena}, H. Fritzche and M. Pollak, eds. (World Scientific,
1990) p.XX.
\item{7.}V. L. Nguyen, B. Z. Spivak, and B. I. Shklovskii, Pis'ma Zh.
Eksp. Teor. Fiz. {\bf 41}, 35 (1985) [Sov. Phys. JETP {\bf 41}, 42
(1985)].
\item{8.}M. Griniasty and S. Fishman, Phys. Rev. Lett {\bf 60}, 1334 (1988).
\item{9.}D. J. Thouless, Phys. Rev. Lett. {\bf 61}, 2144 (1988).
\item{10.}S. Das-Sarma, S. He and X. C. Xie, Phys. Rev. Lett {\bf 61},
2144 (1988); and Phys. Rev. B{\bf 41}, 5544 (1990).
\item{11.}S. Das-Sarma, A. Kobayashi and R. E. Prange, Phys. Rev. Lett {\bf
56}, 1280 (1986); and Phys. Rev. B{\bf 34}, 5309 (1986).
\item{12.}S. Fishman, D. R. Grempel and R. E. Prange, Phys. Rev. Lett. {\bf
49},
509 (1982); D. R. Grempel, R. E. Prange and S. Fishman, Phys. Rev. A{\bf 29},
1639 (1984).
\item{13.}M. V. Berry and J. Goldberg, Nonlinearity {\bf 1}, 1 (1988).
\item{14.}E. A. Coutsias and D. Kazarinoff, Physica {\bf 26D}, 295 (1987).
\item{15.}G. H. Hardy and J. E. Littlewood, Acta Math. {\bf 37}, 25 (1914).
\item{16.}N. Brennan and S. Fishman, Nonlinearity {\bf 5}, 211 (1992).
\item{17.}H. J. St\"ockmann, private communication.
\item{18.}M. V. Berry, Physica {\bf 10D}, 369 (1984).
\item{19.}A. Ya. Khinchin, {\it Continued Fractions} (U. Chicago Press, Chicago
1964).
\item{20.}G. H. Hardy and E. M. Wright {\it An Introduction to the Theory of
Numbers}, 5th edition, (Oxford Press, New York 1979).
\item{21.}P. G. de Gennes, C. R. Acad. Sci. B{\bf 292}, 9, 279 (1981); S.
Alexander, Phys. Rev. B{\bf 27}, 1541 (1983).
\item{22.}B. Pannetier, J. Chaussy and R. Rammal, Jpn. J. of Appl. Phys.
Suppl. {\bf 26}, 1994 (1987).
\item{23.}J. N. Gordon, A. M. Goldman, J. Maps, D. Costello, R. Tiberio,
and B. Whitehead, Phys. Rev. Lett. {\bf 56}, 2280 (1986).
\item{24.}J. Simonin, D. Rodriguez, and A. Lopez, Phys. Rev. Lett. {\bf 49},
944 (1982); H. J. Fink, A. Lopez, and R. Maynard, Phys. Rev. B{\bf 26}, 5237,
(1982).

\vfill\eject

\noindent{\bf Figure Captions}
\vskip 10pt

\item {Fig. 1.}(a) $A_N$ for $\alpha={1\over 2}(\sqrt{5}-1)$,
{(b)} ${\rm max}A_N$ for $\alpha={1\over 2}(\sqrt{5}-1)$,
{(c)} $A_N$ for $\alpha=1/e$,
\break {(d)} ${\rm max}A_N$ for $\alpha=1/e$.
\item{Fig. 2.}The sums $A_N^\nu$ as a function of $x=N^\nu$ for
$\alpha={1\over 2}(\sqrt{5}-1)$ and (a) $\nu=1/3$, (b) $\nu=1/2$, and (c)
$\nu=0.8$.
\item{Fig. 3.}The Fourier transform of the sums of Fig.~2.
\item{Fig. 4.}The Fourier transform of the sequence of Fig.~1(a).
\item{Fig. 5.}The Sums $A_N^\nu$ for $\alpha={1\over 2}(\sqrt{5}-1)$ and
$\nu=1.5$ as a function of $N$.
\item{Fig. 6.}The sums $A_N^\nu$ as a function of $N$ with $\alpha={1\over2}
(\sqrt{5}-1)$
for (a) $\nu=3$ and (b) $\nu=2$.

\vfill\eject

\bye